\pdfoutput=1
\documentclass[11pt]{article}
\usepackage[]{acl}
\usepackage{times}
\usepackage{amsmath}
\usepackage{mdframed}
\usepackage{latexsym}
\usepackage{multirow}
\usepackage{booktabs}
\usepackage{array}
\usepackage{verbatimbox}
\usepackage{multicol}
\usepackage[inline]{enumitem}
\usepackage[T1]{fontenc}
\usepackage[utf8]{inputenc}
\usepackage{microtype}
\usepackage{inconsolata}
\usepackage{graphicx}

\newcommand{\blue}[1]{\textcolor{black}{#1}}
\newcommand{\green}[1]{\textcolor{black}{#1}}
\newcommand{\header}[1]{\vspace*{1mm}\noindent\textbf{#1}.}
\definecolor{verylightgray}{gray}{0.95} %
\title{HotelMatch-LLM: Joint Multi-Task Training of Small and Large Language Models for Efficient Multimodal Hotel Retrieval}
\author{Arian Askari\textsuperscript{1}, Emmanouil Stergiadis\textsuperscript{2}, Ilya Gusev\textsuperscript{2}  \\ \textbf{Moran Beladev\textsuperscript{2}} \\
\textsuperscript{1}Leiden University a.askari@liacs.leidenuniv.nl \\
\textsuperscript{2}Booking.com \{emmanouil.stergiadis,ilya.gusev,moran.beladev\}@Booking.com \\
}
\begin{document}
\maketitle
\begin{abstract}
We present HotelMatch-LLM, a multimodal dense retrieval model for the travel domain that enables natural language property search, addressing the limitations of traditional travel search engines which require users to start with a destination and editing search parameters. HotelMatch-LLM features three key innovations: (1) Domain-specific multi-task optimization with three novel retrieval, visual, and language modeling objectives; (2) Asymmetrical dense retrieval architecture combining a small language model (SLM) for efficient online query processing and a large language model (LLM) for embedding hotel data; and (3) Extensive image processing to handle all property image galleries.
Experiments on four diverse test sets show HotelMatch-LLM significantly outperforms state-of-the-art models, including VISTA and MARVEL. 
Specifically, on the test set—main query type—we achieve 0.681 for HotelMatch-LLM compared to 0.603 for the most effective baseline, MARVEL.
Our analysis highlights the impact of our multi-task optimization, the generalizability of HotelMatch-LLM across LLM architectures, and its scalability for processing large image galleries.
\end{abstract} %

\section{Introduction}
Online property search platforms are essential for modern travel, enabling millions of people to book accommodations with generating $667$ billion in revenue in 2023 \cite{travelperk_online_travel_booking} underscoring their importance. However, current systems\footnote{Examples are Airbnb.com, Booking.com, and Hotels.com, which are widely used for accommodation search and booking.} are often limited to pre-defined filters, where users must first select a country and city before applying additional criteria like price, or star rating. This process limits flexibility, preventing users from expressing preferences in natural language or exploring options without a set destination. For example, searches like "hotels with pantone curtains" or "beachfront accommodations near a park" are unsupported. 
\par
To overcome these limitations, we propose HotelMatch-LLM,\footnote{In this wreview=falseork, the term `hotel' is used as a general reference to various types of accommodations, including but not limited to hotels, bed and breakfasts, and private homes.} a multimodal retrieval model designed to support free-form natural language queries for hotel search. Our model features three novel components to enhance property search: \begin{enumerate*}[label=(\roman*)] \item \textit{Domain-Specific Multi-Task Optimization}: Multi-task optimization tailored to the travel domain, with objectives for retrieval alignment, masked language modeling (MLM) for contextual understanding, and visual facility learning for recognizing amenities in hotel images. \item \textit{Asymmetrical Dense Retriever Architecture}: A novel architecture combining an SLM\footnote{We refer to models of 110 million parameters as SLMs and to language models ranging from 330 million to 7 billion parameters as LLMs.} for efficient online query processing and an LLM for embedding of hotel data. This approach achieves near-LLM performance with SLM efficiency. \item \textit{Multiple Image Processing}: A mean pooling strategy over patch-level embeddings enables processing an extensive, theoretically unlimited, number of images per property, creating a fixed-size representation that captures comprehensive visual context. \end{enumerate*}
\par
\begin{table}[]
\centering
\scalebox{0.85}{
    \begin{tabular}{p{8.5cm}}
    \toprule
    \textbf{Multimodal (main).} private room, pool, mountain view, and a cot available for a baby.                                          \\ \midrule
    \textbf{Text-driven.} Central apartment near Serbian landmarks with quick access to Belgrade airport.                                           \\ \midrule
    \textbf{Vision-driven.} Spacious rooms with white bedding and geometric headboards, modern meeting room with orange chairs, indoor pool, vibrant dining area with leaf-themed decor \\ \midrule
    \textbf{Out-of-distribution.} I'm looking for a 3-star or better hotel next to a subway station, no more than 30 minutes away from Tokyo station. If it's 1 room, I need 2 or 3 beds, and if it's 2 total rooms, then 1 bed in each room. \\ \bottomrule
    \end{tabular}
}\caption{Examples of query types in our dataset.}
\label{tab:query_types}
\end{table}
We train HotelMatch-LLM using synthetic relevance labels generated by GPT-4o given a pair of query and property, building on prior works demonstrating a strong correlation between GPT-4-generated labels and human annotations \cite{SyntheticTestCollectionSIGIR2024,LLMsPredPrefSIGIR2024,upadhyay2024umbrela}. We use the prompt shown in Figure \ref{fig:gen_label} detailed in Section \ref{sec:appendix_gen_label}. To evaluate performance, we use the HotelMatch dataset, which contains 3 million multimodal documents and four distinct test sets. Examples of these test sets are shown in Table \ref{tab:query_types}, with descriptions in Section \ref{sec:dataset}.
\par
Our contributions are as follows: 
\begin{enumerate*}[label=(\roman*)] 
    \item We introduce HotelMatch-LLM, a novel multimodal dense retrieval model designed specifically for the travel domain, enabling natural language questions for property search, overcoming the limitations of existing systems.
    \item Our model outperforms state-of-the-art (SOTA) multimodal retrieval models designed for web search including VISTA and MARVEL models \cite{zhou-etal-2024-vista,zhou-etal-2024-marvel} on four distinct test sets.
    \item We propose a multi-task optimization framework that integrates textual and visual features. This includes retrieval optimization to align query and document embeddings, masked language modeling to enhance contextual understanding, and visual facility learning to identify key amenities from images.
    \item We conducted comprehensive ablation study, evaluating the contribution of each task in our multi-task optimization.%
\end{enumerate*}

\section{Related Work}
\header{Text-Driven Dense Retrievers}
Dense retrieval systems for text-driven search have evolved significantly with the advent of LLMs. Early work on dense retrievers, such as DPR \cite{karpukhin2020dense} and ANCE \cite{xiong2021approximate}, focused on using embeddings to represent queries and documents for fast and accurate retrieval based on cosine similarity. More recent models, such as ColBERT \cite{khattab2020colbert} and GTR \cite{ni2021gtr}, have further optimized dense retrievers for efficiency and scalability. However, these systems often rely on LLMs for both query and document embedding, which can be computationally expensive for real-time search scenarios. To address this, our HotelMatch-LLM adopts a hybrid approach, using an SLM for online query embedding, while utilizing the larger LLM for offline hotel data embeddings.
\par
\header{Multimodal Dense Retrievers}
Multimodal retrieval systems integrate text and visual data to enhance information retrieval. Models like CLIP \cite{radford2021learning} have proven effective for aligning these modalities. Recent advancements, such as MARVEL \cite{zhou-etal-2024-marvel} and VISTA \cite{zhou-etal-2024-vista}, achieve high performance but are limited to processing single images per document. In contrast, our HotelMatch-LLM enables processing multiple images per property, addressing the unique demands of the travel domain.
\begin{figure*}
    \centering
    \scalebox{0.62}{\includegraphics[]{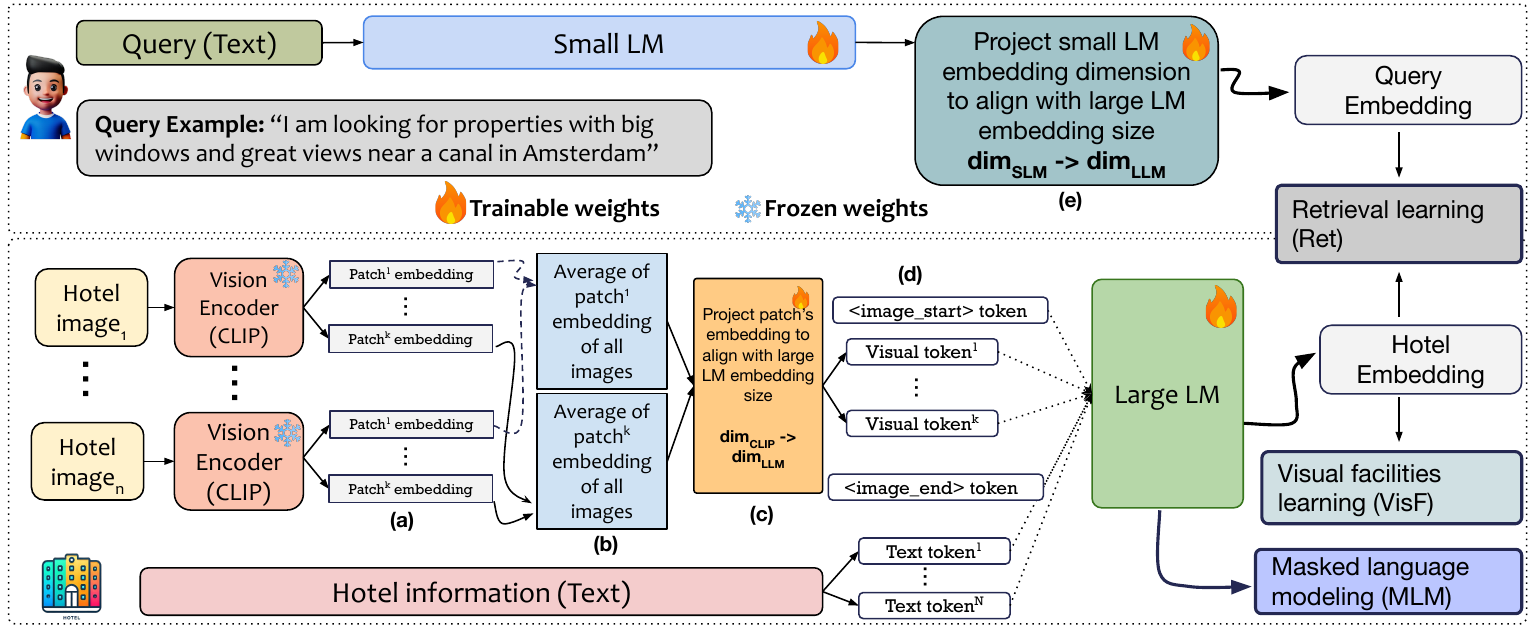}}
    \caption{An illustration of the training setup for the proposed method, HotelMatch-LLM. The \textbf{top part} encodes the query text using a small LM and aligns its embedding with the large LM, while the \textbf{bottom part} processes hotel images using a vision encoder (CLIP) to extract patch embeddings, averages them, aligns them with the large LM embedding size, and jointly passes them through the large LM to obtain the final hotel embeddings. Finally, a multi-task optimization is applied to train the HotelMatch-LLM retriever. $dim_\text{CLIP}$, $dim_\text{SLM}$, and $dim_\text{LLM}$ refer to the embedding dimensions of the CLIP, small LM, and large LM, respectively.}%
    \label{fig:proposed_method}   
\end{figure*}
 
\par
\header{Multimodal LLMs}
LLMs have increasingly been extended to multimodal tasks, enabling simultaneous text and image processing. Prominent examples include LLaVA \cite{liu2023visual}, BLIP-2 \cite{li2023blip}, and Flamingo \cite{alayrac2022flamingo}, which excel in tasks like image captioning and visual question answering. However, our experiments showed that language models pre-fine-tuned for text retrieval achieve significantly faster convergence in optimizing multimodal retrieval within the travel domain. To address the computational challenges of fine-tuning from scratch, we leveraged such pre-fine-tuned models and further fine-tuned them for multimodal retrieval. Consequently, we excluded models like Qwen-VL \cite{bai2023qwen}, LLaVA, and Pixtral \cite{agrawal2024pixtral}, which focus on multimodal generation but lack pre-fine-tuning for text retrieval.
\blue{}

\section{Proposed Method: HotelMatch-LLM}
\blue{We tackle the task of multimodal retrieval in the travel domain with our proposed method, HotelMatch-LLM. Given a query $q$, the task involves using a dense retrieval model to search for relevant documents from a collection  $D$ to address the user’s information needs \cite{zhou-etal-2024-marvel,xiong2021approximate}. We detail the components of HotelMatch-LLM in the following.}
\subsection{Extensive Number of Images} 
\blue{A key feature of HotelMatch-LLM is its ability to process a complete accommodation gallery in the form of an extensive number of images alongside the textual content of each accommodation.}
Unlike existing multimodal models designed for web search \cite{zhou-etal-2024-vista,zhou-etal-2024-marvel} that are restricted to single-image inputs, HotelMatch-LLM employs an effective method to aggregate information from all images, creating comprehensive embeddings. 
\blue{We formally elaborate below on this feature by starting with image-level embedding.}
\par
\header{Image-Level Embedding} 
Each image $j$ in a document, denoted as $d_{\operatorname{img}_j}$, is processed through the CLIP encoder to obtain an image-level embedding. Formally:  
\begin{equation}
    \textbf{h}_{\operatorname{img}_j} = \operatorname{CLIP}(d_{\operatorname{img}_j}),
\end{equation}
where $\textbf{h}_{\operatorname{img}_j}$ represents the embedding for the $j$-th image in the document\blue{, and is divided into $k$ patches, shown as step (a) in Figure \ref{fig:proposed_method}, generating $k$ patch embeddings}. This representation is obtained by processing the $j$-th image through the CLIP visual encoder, specifically leveraging the grid features extracted from the last layer of the CLIP visual encoder\blue{,}
generating $k$ patch embeddings:
\begin{equation}
    \textbf{h}_{\operatorname{img}_j} = \{\textbf{h}^{\mathrm{1}}, \textbf{h}^{\mathrm{2}}, \dots, \textbf{h}^{\mathrm{k}}\}.
\end{equation}
We set $k$ to $49$ as we resize all images to $224\times224$ pixels cropped to the center, and use the CLIP model with a $32\times32$ window size.\footnote{For more details on the relationship between image resolution and patch count, we refer readers to \cite{radford2021learning}.}
\par
\header{Mean Pooling Across Images}
To handle an extensive, theoretically unlimited, number of images, a mean pooling operation is applied across the corresponding patches of all images, as shown in step (b) in Figure \ref{fig:proposed_method}.
Assuming a hotel with $N$ images, each with $k$ patch embeddings, pooling produces a tensor of dimensions $(k \times \operatorname{embedding\,dimension})$. The mean pooling is performed as follows:
\begin{equation}
    \textbf{h}^{i}_{\mathrm{pooled}} = \frac{1}{N} \sum_{j=1}^{N} \textbf{h}^{i}_{\operatorname{\operatorname{img}}_j},
\end{equation}
where $ \textbf{h}^{i}_{\mathrm{pooled}}$ is the pooled embedding for the $i$-th patch across all images in the document, \blue{and $\textbf{h}^{i}_{\operatorname{\operatorname{img}}_j}$ 
denotes the embedding of the $i$-th patch for the $j$-th image of document $d$.}
\par
\header{Fixed-Size Representation} The final representation for all images associated  with a document is represented in $k$ vectors and formed by concatenating these pooled embeddings:
\begin{equation}
    \textbf{h}_d = \{\textbf{h}^{\mathrm{1}}_{\mathrm{pooled}}, \textbf{h}^{\mathrm{2}}_{\mathrm{pooled}}, \dots, \textbf{h}^{\mathrm{k}}_{\mathrm{pooled}}\},
\end{equation}
where \blue{$ \textbf{h}^{i}_{\mathrm{pooled}}$} represents the pooled embedding for the $i-th$ patch across all $N$ images associated with document $d$. This operation is repeated for each of the $k$ patches ($k=49$), resulting in a representation of size $(k \times \operatorname{embedding\,dimension})$. This fixed size ensures scalability for any number of images, as the dimensions remain constant regardless of $N$.\footnote{\blue{We acknowledge that some information loss might occur due to pooling. We explore alternative options in Section \ref{sec:results}, and find the aforementioned approach to be the most effective.}} %

\par
\header{Projection into Textual Space} \blue{The fixed-size pooled representation is} then projected into the same space as textual tokens of the pre-trained language model via a dense linear transformation, shown in step (c) in Figure \ref{fig:proposed_method}:
\begin{equation}\label{eq:project_img_to_text_vec_space}
    \textbf{I}^{i}_{\mathrm{pooled}} = \operatorname{linear}( \textbf{h}^{i}_{\mathrm{pooled}}),
\end{equation}
producing embeddings $\textbf{I}_{\operatorname{\mathrm{pooled}, i}}$ which can be concatenated with textual embeddings to form the final document representation since the input dimension of linear is embedding dimension of visual encoder of CLIP and the output dimension of linear layer is the dimension size of textual embedding of language model, shown in step (d) in Figure \ref{fig:proposed_method}:
\begin{multline} \label{eq:input_embedding_vector}
  \begin{gathered}
    \textbf{X} = \textbf{e}(<\operatorname{start}>) ;\textbf{I}_{\mathrm{pooled}}^{1}; \dots; \textbf{I}_{\mathrm{pooled}}^{k} \\
    ;\textbf{e}(<\operatorname{end}>); \textbf{e}^{1}; \dots; \textbf{e}^{M},
    \end{gathered}
\end{multline}
Here, \blue{$\textbf{X}$ is the final input representation of the property,} $;$ denotes concatenation operation, and $\textbf{e}(<start>)$ and $\textbf{e}(<end>)$ are the embedding of visual separator tokens that mark the start and end of the image feature representations. $\{ \textbf{e}^{1}, \dots, \textbf{e}^{M}\}$ 
are the token embeddings of the text input sequence the hotel’s textual description. 
\green{Finally, we pass $\textbf{X}$ through the LLM to make the final representation of the property:}
\begin{equation} \label{eq:embedding_vector}
    \textbf{d} = LLM(\textbf{X})
\end{equation}
where $\textbf{d}$ is representation of the property which is either the representation of the CLS token or the mean pooling over the input tokens, depending on the strategy suggested by the language model (LM) used as the backbone.
This design integrates both visual and textual signal into the model and enables the representation of an unlimited number of images as $k$ visual tokens, in contrast to previous studies, which support only a single image. To provide more insights, Figure \ref{fig:input_structure_text}, presented in the Appendix section, illustrates a detailed example of how the property information is structured and prompted to HotelMatch-LLM.

\subsection{Joint training of SLM and LLM}
\blue{LLMs demonstrate strong effectiveness as embedders \cite{lee2024nv,wang-etal-2024-improving-text,li2024makingtextembeddersfewshot}, and ranked as the top-10 most effective models in well-known benchmarks such as MTEB \cite{muennighoff-etal-2023-mteb}. However, deploying LLMs for query embedding at inference time introduces significant computational overhead, making it a challenging and active area of research \cite{park2024anyprecision}. These challenges are particularly significant in property search systems, where millions of users submit queries daily, necessitating both efficiency and cost-effectiveness in embedding models. 
To overcome this, we introduce a novel asymmetrical architecture in HotelMatch-LLM. We use an LLM as the backbone for embedding documents (bottom part of Figure \ref{fig:proposed_method}), while we use an SLM backbone for embedding queries (top part of Figure \ref{fig:proposed_method}).
As the embeddings of the query, denoted as $\textbf{q}$, and document, denoted as $\textbf{d}$, should have the same dimensional size to compute cosine similarity, we add a dense linear layer that projects the SLM's embedding dimension, shown in step (e) in Figure \ref{fig:proposed_method}, to match the LLM's embedding dimension.
Our experiments show that this setup is more effective than using SLM for embedding both queries and documents, while maintaining the same efficiency at inference time as using SLM for embedding both queries and documents at query time. This could be attributed to the greater complexity inherent in hotel data compared to query data, allowing an LLM to represent it more effectively in vector space than an SLM.
}
\blue{We found the most important factor in this joint training is how learning rates (LR) are set. We apply distinct learning rates: a higher rate for the SLM and a lower rate for the LLM. This choice is made by our empirical observations and aligns with findings from previous research indicating that larger language models benefit from lower learning rates, whereas smaller models perform better with higher rates \cite{kaplan2020scaling}. Further analysis of this approach is provided in Section \ref{sec:joint_training}.}
\subsection{Domain-specific Multi-task Optimization}
\blue{In the travel domain, geography (city and country) and facilities are key features for businesses, often mentioned in hotel descriptions or visible in property images. Our multi-task loss is designed to capture these features effectively. The visual facility learning (VisF) loss focuses on identifying facilities from property embeddings (denoted as $\textbf{d}$ in Equation \ref{eq:embedding_vector}). The labels for this task are collected by passing all the images of a property to the MUMIC method \cite{mumicAAI2023} in order to identify facilities from property's images. We follow MUMIC methodology and the list of 120  facility labels in the MUMIC paper. These labels include visually identifiable features such as swimming pools, gyms, and balconies. Meanwhile, the MLM loss predicts masked city and country tokens in descriptions, ensuring a strong textual understanding of geographic features. We next formally define our primary objective, retrieval learning, and introduce the two domain-specific losses.}
\par
\header{Retrieval learning}
The primary objective is to optimize the model for distinguishing relevant documents (hotels) from irrelevant ones using a contrastive learning approach. Following \cite{zhou-etal-2024-marvel}, for a query embedding $\textbf{q}$,
a positive document $d^+$, and a set of negative documents
$\{d^-_1, \dots, d^-_n\}$, 
the similarity score is defined as:
\begin{equation}
    S(q, d) = \operatorname{cosine}(\textbf{q}, \textbf{d}),
\end{equation}
The probability of \( d^+ \) being relevant to \( q \) is computed \blue{using the softmax function over the cosine scores, as follows}:
\begin{equation}
    P(d^+|q) = \frac{\exp(S(q, d^+))}{\sum_{d \in \{d^+, d^-_1, \dots, d^-_n\}} \exp(S(q, d))}.
\end{equation}
The retrieval loss, formulated as cross-entropy, is expressed as:
\begin{equation}
    \mathcal{L}_{\operatorname{Ret}} = -\log P(d^+|q).
\end{equation}
\par
\header{Masked Language Modeling}
Let \( t = \{t_1, \dots, t_N\} \) represent the \blue{textual} tokens of a document. After masking the city and country tokens, the model predicts logits \( \hat{t}_i \) for these masked positions, next the probability is computed using the softmax function, applied to the output of the model's MLM head:
\begin{equation} \small
   P(t_i \mid h_i^{\operatorname{masked}}) = \frac{\exp(\operatorname{MLM\_Head}(h_i^{\operatorname{masked}})[t_i])}{\sum_{j=1}^{|\mathcal{V}|} \exp(\operatorname{MLM\_Head}(h_i^{\operatorname{masked}})[j])}
\end{equation}
where $P(t_i \mid h_i^{\operatorname{masked}})$ is the conditional probability of the token $t_i$, given the masked token's hidden state 
 \( h_i^{\operatorname{masked}} \), and $|\mathcal{V}|$ refers to the size of the vocabulary.
The MLM\_Head is a linear layer where the input has the same dimension as the token embeddings of the large language model, and the output dimension corresponds to the size of the vocabulary.
The loss is computed as:
\begin{equation}
    \mathcal{L}_{\operatorname{MLM}} = -\frac{1}{T} \sum_{i=1}^T \log P(t_i | h_i^{\operatorname{masked}}),
\end{equation}
where \( T \) is the total number of masked tokens. \blue{In our experiment, \( T \) is dynamic and depends on the number of tokens required to represent the country and city of the hotel.} %
\par
\header{Visual Facility Learning}
We optimize the model to predict the presence of facilities (e.g., pool, gym) using hotel embeddings that are represented as \( \textbf{d} \). A linear layer with \(\mathcal{F}\) outputs (where \(\mathcal{F}\) is the total number of facilities whose presence is identified using the MUMIC method \cite{mumicAAI2023}, and \(\mathcal{F} = 120\), followed by a sigmoid activation function, computes the probability of each facility:
\begin{equation}
    \hat{f}_i = \sigma(W_f \cdot \textbf{d} + b_f),
\end{equation}
where \( \textbf{d} \) is the document embedding, and \( W_f \) and \( b_f \) are the learnable weights and biases. The binary cross-entropy loss for this task is:
\begin{equation}
    \mathcal{L}_{\operatorname{VisF}} = -\frac{1}{\mathcal{F}} \sum_{i=1}^{\mathcal{F}} \big[f_i \log \hat{f}_i + (1 - f_i) \log (1 - \hat{f}_i)\big],
\end{equation}
where \( f_i \) is the ground-truth facility label. %
\par
\header{Final Loss Aggregation}
The final loss combines the three objectives, weighted by empirically determined values (\( \lambda_1 = 0.7 \), \( \lambda_2 = 0.2 \), and \( \lambda_3 = 0.1 \)). These weights were optimized through a grid search on the validation set, testing values in increments of 0.1 to achieve the best retrieval performance:
\begin{equation}
    \mathcal{L}_{\operatorname{final}} = \lambda_1 \mathcal{L}_{\operatorname{Ret}} + \lambda_2 \mathcal{L}_{\operatorname{MLM}} + \lambda_3 \mathcal{L}_{\operatorname{VisF}}.
\end{equation}

\begin{table}[]
\centering
\caption{Dataset statistics of our dataset, HotelMatch. `Q' and `D' refer to query and hotel.}
\label{tab:data_stats}
\begin{tabular}{lll} 
\toprule    
Number of  Documents (Hotels)               & 3.1M          \\ \midrule
Avg number of images per D  & 44.6                  \\ 
Avg number of words per D   & 185.9                   \\ 
Avg number of words per Q   & 5.4                 \\ \midrule
Number of Train Queries           & 57,884               \\
Number of Validation Queries      & 500                   \\
Number of Test Queries        & 1000                  \\ 
\bottomrule
\end{tabular}
\end{table}
\begin{table*}[ht]
\centering
\caption{The results of our model compared to the baselines. MRR and nDCG refer to MRR@10 and nDCG@10. Significance is shown with $\dagger$ for the best result (HotelMatch-LLM) compared to most effective baseline, MARVEL. Statistical significance was measured with a paired t-test ($p<0.05$) with Bonferroni correction for multiple testing.}
\label{tab:main_results}
\scalebox{1}{\begin{tabular}{l|c|c|c|c|c|c|c|c} 
\toprule
Test Query Set Name & \multicolumn{2}{c|}{Real-world}         & \multicolumn{2}{c|}{Vision-driven}   & \multicolumn{2}{c|}{Text-driven}  & \multicolumn{2}{c}{Out-of-distribution}    \\ \midrule
Size                     & \multicolumn{2}{c|}{1000 queries}       & \multicolumn{2}{c|}{101 queries}        & \multicolumn{2}{c|}{101 queries}  & \multicolumn{2}{c}{100 queries}    \\ \midrule
Model                         & MRR        & nDCG & MRR@            & nDCG & MRR   & nDCG  & MRR   & nDCG \\ \midrule
\multicolumn{9}{c}{Setting: Text-only Modality} \\ \midrule
BM25                     & .506             & .401      & .138                  & .195       &        .798          & .825  & .588 & .489   \\
CLIP-Text (Zero-Shot)    & .452             & .381       & .140                  & .197       &         .541         & .600     & .559 & .428     \\
GTR-base  (Zero-Shot)   & .547             & .426       & .142                  & .219      &         .812        & .843    & .650 & .521      \\
GTR-large (Zero-Shot)   & .545             & .429       & .148                  & .224       &           .824       & .857   & .656 & .534    \\ \midrule
\multicolumn{9}{c}{Setting: Multimodal (Image and text)}          \\ \midrule
CLIP     (Zero-Shot)     & .460             & .402       & .172                  & .254      &       .545           & .609      & .561 & .439    \\
MARVEL (Fine-tuned) & .603             & .503       & .219                  & .326       &          .810       & .833     & .660 & .515    \\
VISTA     (Fine-tuned)   & .582             & .465       & .216                 & .321       &       .802           & .839     & .662 & .513     \\
HotelMatch-LLM (Ours)         & \textbf{.681$^\dagger$}             & \textbf{.600$^\dagger$}       & \textbf{.247$^\dagger$}                 & \textbf{.362$^\dagger$}       &       \textbf{.863$^\dagger$}          & \textbf{.884$^\dagger$}     & \textbf{.704$^\dagger$} & \textbf{.558$^\dagger$}     \\ 
\midrule
HotelMatch-LLM w/o vision         & .595             & .482       & .154                  & .239       &      .829    &      .863            & .658 &   .535    \\
\bottomrule
\end{tabular}}
\end{table*}
\section{Experimental Setup}
\header{Dataset}\label{sec:dataset}
For our experiments, we utilize our HotelMatch dataset, designed for multimodal hotel retrieval. Table \ref{tab:data_stats} summarizes its statistics.\footnote{Our test set, including top-100 candidate documents' links considered for each query along with their labels, will be released to foster research in this area.} %
Queries are categorized to four distinct query types (examples in Table \ref{tab:query_types}): \begin{enumerate*}[label=(\arabic*)] \item Multimodal Queries: User queries have been synthesized making sure production distribution is preserved, by utilizing an LLM to rephrase anonymized queries, \item Vision-driven Queries: Synthesized from hotel images using the prompt illustrated in Figure \ref{fig:vision_driven_qgen}, \item Text-driven Queries: Synthesized from property descriptions using the prompt illustrated in Figure \ref{fig:text_driven_qgen}, \item 
Out-of-distribution Queries: Travel queries from a significantly different distribution, see Table \ref{tab:query_types}. %
\end{enumerate*} 
\par
\noindent
\header{Evaluation}
Our experiments focus on re-ranking 100 candidates retrieved by a fine-tuned CLIP-based model to manage computational costs. Irrelevant queries are excluded. The best-performing re-ranking models are extended to full-ranking, as shown in Tables \ref{tab:main_results} and \ref{tab:full_retrieval}. We utilize GPT-4o for synthetic annotations of top-100 candidates, using the prompt shown in Figure \ref{fig:gen_label} detailed in Section \ref{sec:appendix_gen_label}, reducing reliance on costly human labeling while maintaining high quality, and performance is evaluated using MRR and nDCG at top-10 results.
\par
\header{Baselines}
We compare HotelMatch-LLM against SOTA multimodal retrieval models, including:
\begin{enumerate*}[label=(\roman*)]
    \item \textbf{MARVEL}: A leading multimodal retrieval model optimized for web search.
    \item \textbf{VISTA}: Another SOTA model known for its effectiveness in various multimodal tasks.
    \item \textbf{GTR-base} and \textbf{GTR-large} \cite{ni2021gtr}: Fine-tuned models specifically adapted for retrieval task.
\end{enumerate*}
We also include unimodal baselines including \textbf{Best Match 25 (BM25}) \cite{robertson1994some} and \textbf{CLIP}. 
\par
\header{Implementation details} 
We implement HotelMatch-LLM in PyTorch \cite{paszke2017automatic}. In all of our training experiments, we fine-tune for 10 epochs with early stopping after five validation steps without improvement. We employ FAISS \cite{douze2024faiss} for efficient k-Nearest Neighbor (KNN) retrieval.
We generate embeddings following the methods recommended by their respective LM backbones (e.g., CLS token or mean pooling over input tokens). We use GTR-Base-110M as the SLM in all our experiments. For the LLM, we use GTR-Large-335M in the main experiments due to its competitive performance and fast convergence. To assess generalizability, we tested larger LMs, including Zeta-Alpha-E5-Mistral and Stella-en, with 7 billion and 1.5 billion parameters.%
We use learning rates (LR) as reported for MARVEL and VISTA, finding optimal LRs of 5e-4 for GTR-base and 5e-6 for GTR-large, Zeta-Alpha-E5-Mistral-7B~\cite{zeta_alpha_e5_mistral} and Stella-en-1.5B \cite{dunzhang_stella_en_1_5B_v5} where they are used as LM backbone in HotelMatch-LLM.
\section{Results} \label{sec:results}
In this section, we answer the following research questions, evaluating the effectiveness of our proposed method, HotelMatch-LLM, from different perspectives:
\begin{itemize}[leftmargin=*,nosep]
    \item \textbf{RQ1:} What is the effectiveness of HotelMatch-LLM compared to existing SOTA multimodal retrievers? %
    \item \textbf{RQ2:} What is the most optimal method to represent a long-image context compared to single-image processing?  %
    \item \textbf{RQ3:} What is the impact of Multitask Optimization in HotelMatch-LLM and what is the importance of each task?
    \item \textbf{RQ4:} How well does our proposed method generalize to other LLM architectures?
\end{itemize}
\begin{table}[ht]
\small
\centering
\caption{Analyzing impact of our approach for representing extensive number of images compared to alternative options. Our approach allows for an unlimited number of images in theory. In practice, the maximum number of images per property in our dataset is 306 images. 1TPI refers to "One Token per Image".}
\label{tab:img_rep}
\begin{tabular}{l|l|c|c}
    \toprule
    Method             & Number of image    & MRR & nDCG  \\
    \midrule
    MARVEL            & Single             & .603   & .503     \\
    \midrule
    \multicolumn{4}{l}{Methods for processing multiple images (Ours)}  \\
    \midrule
    HotelMatch  & Multiple-unlimited & .681   & .600     \\ \midrule
    1TPI-Patch & Multiple-limited   & .672   & .585     \\ \midrule 
    1TPI-CLS   & Multiple-limited   & .652  & .580    \\ 
    \bottomrule
\end{tabular}
\end{table}
\begin{table}[ht]
\small
\centering
\caption{Generalizability of HotelMatch-LLM, using different models as LLM backbone.}
\label{tab:gen_to_other_llms}
\begin{tabular}{l|p{2.1cm}|c|c}
    \toprule
    Query embedder                      & Hotel embedder                      & MRR  & nDCG  \\  \midrule
    \multicolumn{2}{c|}{GTR-Large-335M} & .687 & .605  \\   \midrule
    GTR-Base-110M            & GTR-Large-335M           & .681 & .600  \\ 
    GTR-Base-110M            & Stella-en-1.5B           & .694 & .619  \\
    GTR-Base-110M            & Zeta-Alpha-E5-Mistral-7B & .719 & .631  \\  
    \midrule
    \multicolumn{2}{c|}{GTR-Base-110M} & .649 & .568 \\  
    \bottomrule
\end{tabular}
\end{table}
\par
\header{Main results (RQ1)} 
The results presented in Table~\ref{tab:main_results} demonstrate that our proposed HotelMatch-LLM model significantly outperforms all baseline models in all four test sets, including the previous SOTA multimodal dense retrievers MARVEL and VISTA models, in both text-only and multimodal settings; showcasing its superior capability to re-rank tasks for accommodation search. 
Additionally, when we ablate the vision component in HotelMatch-LLM (i.e., HotelMatch-LLM without vision), the performance notably decreases, highlighting the importance of multimodal integration in achieving optimal results. Furthermore, when focusing on unimodal baselines, text-driven GTR models and vision-driven CLIP variants fall short compared to our multimodal approach. %
\par
\header{Extensive number of images (RQ2)} 
To assess our method for representing an extensive number of images in greater depth, we propose two alternative methods for representing multiple images, both utilizing the CLIP encoder to generate image-level representations. Each image is transformed into a single embedding vector, which is then projected into the textual space of the LM using a dense linear layer. Since each image corresponds to a single embedding, it occupies one token in the language model’s input. This approach enables the representation of a limited number of images, constrained by the language model's maximum input token capacity minus the tokens reserved for textual descriptions of properties. The first method, 1TPI-CLS, leverages the CLS token embedding from CLIP to represent the entire image. This representation is projected to align with the language model's token dimension as a visual token. The second method, 1TPI-Patch, aggregates information by averaging the patch embeddings produced by CLIP and projects the resulting aggregated embedding as a visual token. We pass a maximum of 50 images, i.e., 50 visual tokens, in our experiments for the proposed alternative methods: 1TPI-Patch and 1TPI-CLS. Table \ref{tab:img_rep} shows that our HotelMatch-LLM method archives the highest effectiveness compared to the SOTA baseline, MARVEL, and alternative methods.
\par
\header{Generalizability to other LLMS (RQ3)} Table \ref{tab:gen_to_other_llms} demonstrates the generalizability of our proposed HotelMatch-LLM method across various LM architectures for query and hotel embeddings. Notably, using the GTR-base-110M model for both the query and hotel embedding yields the lowest effectiveness, suggesting that relying solely on a smaller model for both tasks limits the model's ability to capture complex hotel information. However, when GTR-base-110M is employed as the query embedder and combined with larger models for hotel embeddings, such as Zeta-Alpha-E5-Mistral-7B or Stella-en-1.5B, the effectiveness improves significantly. The best results are achieved when using GTR-base-110M for query embeddings and Zeta-Alpha-E5-Mistral-7B for hotel embeddings, which provides the highest MRR and nDCG scores. This setup maintains the efficiency of GTR-base at query \blue{inference} time while leveraging the more expressive capacity of larger models for encoding hotel information. This pattern suggests that larger language models are better suited for representing the more complex and diverse attributes of hotels, highlighting the scalability and generalizability of the HotelMatch-LLM approach across different architectures.
\par
\begin{table}
\small
\centering
\caption{Analyzing importance of each task in our multi-task optimization.}%
\label{tab:ablation_losses}
\begin{tabular}{l|c|c}
\toprule
Model                               & MRR@10  & nDCG@10  \\ \midrule
HotelMatch-LLM (Ours)                    &      &       \\ \midrule
\textbf{Full model}  & \textbf{.681}  & \textbf{.600} \\ \midrule
$\,$ w/o VisF loss                         & .664 & .575  \\
$\,$ w/o MLM  & .650 & .568  \\ %
$\,$ w/o VisF and MLM             & .632 & .552  \\
\bottomrule
\end{tabular}
\end{table}
\header{Impact of multi-task optimization}
Table \ref{tab:ablation_losses} presents the results of our ablation study, assessing the importance of each task in our multi-task optimization. By systematically removing losses, we evaluate their effects on model effectiveness. The full model achieves the highest effectiveness, while removing MLM or VisF causes declines, highlighting their complementary roles. The most significant drop occurs when MLM is removed, showcasing that geographical understanding has a greater impact on overall effectiveness.
\section{Discussion}
\begin{table}
\centering
\caption{Results of full-ranking. Comparative analysis of our model, HotelMatch-LLM, and the most effective baselines, MARVEL.}
\label{tab:full_retrieval}
\begin{tabular}{l|c|c}
\toprule
          & MRR@10 & nDCG@10  \\ \midrule
HotelMatch-LLM &    .675    &   .592       \\ \midrule
MARVEL    &    .589    & .498 \\
\bottomrule 
\end{tabular}
\end{table}
\header{Full-ranking} Although this paper primarily focuses on re-ranking the top-100 results generated by the initial retriever, where the initial top-k retrieved items are ranked based on relevance and conversion, we also investigate the full-retrieval potential of our method. To achieve this, we evaluate the performance of our proposed method, HotelMatch-LLM, against the most effective baseline, MARVEL, in a full-ranking context \blue{where $3.1$ Million documents are ranked given each query}. To this end, we annotated the top-10 documents retrieved by each model using GPT-4o. Table \ref{tab:full_retrieval} indicates that HotelMatch-LLM outperforms MARVEL in the full-ranking setup.
\par
\begin{table}
\centering
\caption{Latency of our model vs. top-2 most effective baselines. `ms' refers to milliseconds.}
\label{tab:latency}
\scalebox{0.90}{
\begin{tabular}{p{3cm}|l|c|c}
    \toprule
    Model     & Latency (ms)    & MRR  & nDCG  \\ \midrule
    VISTA     & 16.17 ± 0.33  & .572 & .465 \\ \midrule
    MARVEL    & 31.07 ± 0.27 & .603 & .503  \\ \midrule
    HotelMatch-LLM & 18.69 ± 0.38 & .681 & .600  \\ \midrule
    HotelMatch-LLM w/o SLM & 25.37 ± 0.34 &  .687 & .605  \\ 
    \bottomrule
\end{tabular}
}
\end{table}
\header{Efficiency} While the efficiency of our method is evident due to its architectural design, we measure the efficiency of it compared to VISTA, MARVEL, and HotelMatch without SLM, where the LLM backbone of HotelMatch-LLM embeds both query and document. GTR-Base-110M and GTR-Large-335 are SLM and LLM in this experiment. Table \ref{tab:latency} shows that while our efficiency is comparable to VISTA, it is two times higher than MARVEL and $1.4$ times higher than using the LLM alone. The SLM for the experiment with HotelMatch w/o SLM is GTR-Large with 333M parameters. This demonstrates that our method not only achieves higher effectiveness but also efficient.
\par
\begin{table}
\small
\centering
\caption{Results of joint training for the SLM query embeddings and LLM hotel embeddings with various learning rate configurations. The optimal learning rate is reported, determined by training for 100 steps and selecting the rate that achieved the highest effectiveness on the validation set. The SLM and LLM backbones for this experiment are GTR-base and GTR-Large.}
\label{tab:separate_LR}
\scalebox{.95}{
    \begin{tabular}{p{2cm}|c|c|c|c}
        \toprule
        LR tuned for           & SLM LR & LLM LR & MRR  & nDCG  \\ \midrule
        SLM                    & \multicolumn{2}{c|}{5e-4}    & .278 & .280  \\ \midrule
        LLM                    & \multicolumn{2}{c|}{5e-6} & .289 & .302  \\ \midrule
        SLM and LLM (Shared-LR)& \multicolumn{2}{c|}{1e-3} & .315 & .339  \\ \midrule
        SLM and LLM (Separate-LR)& 5e-4   & 5e-6   & \textbf{.681} & \textbf{.600}  \\
        \bottomrule
    \end{tabular}
}
\end{table}
\header{Asymmetrical architecture} \label{sec:joint_training}
Training dense retrieval models with separate encoders for query and document embeddings poses unique challenges. While the theoretical framework for joint training appears straightforward, we found that practical implementation reveals complexities that require careful consideration. 
Table \ref{tab:separate_LR} highlights the impact of different learning rate configurations on model effectiveness. Learning rates were systematically adjusted over 100 training steps, after which the configurations that yielded the highest effectiveness on the validation set were identified. This tuning process highlights the significance of using separate learning rates for the SLM for query embeddings and the LLM for hotel embeddings.%
\par
\section{Conclusions}
We introduced HotelMatch-LLM, a multimodal dense retrieval model for the travel domain. It enables natural language hotel searches, overcoming traditional filter-based limitations. Our multi-task optimization captures structured text and visual attributes from hotel images. By integrating an SLM for query processing and an LLM for embedding complex hotel data, the model delivers near-LLM performance with improved efficiency.
Extensive evaluation on four test sets show HotelMatch-LLM surpasses SOTA models.
Overall, HotelMatch-LLM represents a significant advancement in multimodal hotel retrieval systems, by allowing for contextually rich searches.
While our study focuses on the travel domain, the challenges addressed are broadly applicable on multimodal corpora with extensive number of images.%
\clearpage
\section{Limitations}
Despite the promising results and contributions of HotelMatch-LLM, several limitations warrant discussion. Firstly, while the model demonstrates superior performance in retrieval tasks, it relies heavily on the quality of the synthetic data generated by GPT-4o. If the annotations contain biases or inaccuracies, this could adversely affect the model's learning process and its subsequent performance.
Furthermore, the effectiveness of HotelMatch-LLM in real-world applications may be influenced by factors not accounted for during training, such as dynamic changes in hotel attributes or user preferences. Future work could explore adaptive learning techniques that allow the model to continuously update and refine its embeddings based on user feedback and evolving data.
Additionally, while HotelMatch-LLM is designed to handle complex natural language queries, it is not trained to support multimodal queries where the query can have both text and image. The model may struggle to interpret such queries effectively, leading to suboptimal retrieval results.
Finally, the current version of HotelMatch-LLM does not incorporate user personalization, which could enhance retrieval effectiveness by tailoring results based on individual user preferences, past interactions, or contextual factors. Integrating personalization mechanisms \cite{liu2020personalization} could significantly improve user satisfaction and relevance of the search results.

\bibliography{custom}

\clearpage
\appendix

\section{Appendix}
\subsection{Prompts}
This section outlines the prompts used for synthetic query generation and categorized by their specific applications.
\subsubsection{Vision-Driven Queries}
The prompt for generating vision-driven queries is illustrated in Figure \ref{fig:vision_driven_qgen}. This approach involves attaching 20 randomly sampled images of the hotel alongside the prompt text, which are collectively passed to GPT-4o. The generated queries encapsulate key visual features, such as room layout, unique design elements, furniture styles, window structures, and overall decor. An example of a query produced using this prompt is provided in Table \ref{tab:query_types}.

\begin{figure}[]
    \begin{mdframed}[backgroundcolor=verylightgray,roundcorner=10pt]
    Generate a concise, natural search query that describes the images, listing key features such as bed setup, unique design elements (like geometric headboards), furniture, window style, and overall decor.
    \end{mdframed}
    \centering
    \caption{The prompt for synthetical generating vision-driven queries where the source of query is images of the hotel.}
    \label{fig:vision_driven_qgen}
\end{figure}

\subsubsection{Text-driven queries}
The prompt for text-driven query generation is shown in Figure \ref{fig:text_driven_qgen}. This prompt was adapted from \cite{chaudhary-etal-2024-relative}, maintaining its original structure to avoid introducing biases in the generated queries. The queries are designed to be contextually specific and highlight the unique features of a hotel without explicitly mentioning its name. An example of a query generated with this prompt can be found in Table \ref{tab:query_types}.

\begin{figure}[]
    \begin{mdframed}[backgroundcolor=verylightgray,roundcorner=10pt]
    Given a hotel description from Booking.com, generate a search query for which the hotel description can be a perfect hotel. 
    Generate a query that is distinct and contextually specific, avoiding unintended matches with other hotels in the dataset. The query should highlight unique attributes of the target hotel without including the hotel name. The query must fit semantically with the description but should not have much lexical word overlap.
    
    A general example:
    description: Premature Ventricular Contractions (PVCs, PVC) Medical Definition of Cardiac stress testing, exercise. Cardiac stress testing, exercise: The exercise cardiac stress testing (EST) is the most widely used cardiac (heart) screening test. The patient exercises on a treadmill according to a standardized protocol, with progressive increases in the speed and elevation of the treadmill (typically changing at three-minute intervals).
    query: what is cardiac testing in medical terms

    description: \{DESCRIPTION\} \\
    query:

    \end{mdframed}
    \centering
    \caption{The prompt for synthetical generating text-driven queries where the source of query is textual content of the hotel.}
    \label{fig:text_driven_qgen}
\end{figure}

\subsection{Generating binary relevance judgement} \label{sec:appendix_gen_label}
\blue{Figure \ref{fig:gen_label} illustrates the prompt used for generating binary relevance labels based on a query, the hotel's textual content, and the facilities detected in the hotel's images by MUMIC, represented as text. The detailed facilities identified by the MUMIC tags eliminates the need to pass all hotel images. To validate this, we compared annotations for 100 queries by using the hotel's actual images versus using the facilities identified by MUMIC. We found a strong correlation, with a Pearson correlation coefficient of 0.95.}

\begin{figure}[]
    \begin{mdframed}[backgroundcolor=verylightgray,roundcorner=10pt]
        You are a search quality rater evaluating the relevance of hotel descriptions and images. Given a query and a hotel's description and images, you must provide a score on an integer scale of 0 or 1 with the following meanings:
        
        \begin{itemize}
            \item 1 = Relevant: The hotel description and images are directly related to the query and provide the information the user is seeking.
            \item 0 = Irrelevant: The hotel description and images do not address the query or provide the necessary information.
        \end{itemize}

        Assume that you are providing a recommendation to the user based on their query. If the hotel description and images are primarily about the query, or contain essential information the user is looking for, mark it 1. Otherwise, mark it 0.\\

        A person has typed [{query}] into a search engine.  
        Result: Consider the following hotel description and images.  \\

        —BEGIN Hotel Name, DESCRIPTION AND IMAGES CONTENT—  \\
        \{HOTEL\_Information\} \\
        —END Hotel Name, DESCRIPTION AND IMAGES CONTENT—  \\

        Instructions:  
        Consider the underlying intent of the search, and decide on a final score of the relevancy of the query to the hotel description and images given the context.  
        Score:
    \end{mdframed}
    \centering
    \caption{The prompt for generating binary relevance label given a query and property.}
    \label{fig:gen_label}
\end{figure}

\subsubsection{Domain-specific Multi-task Optimization}
\subsubsection{A detailed example of input format}
Figure \ref{fig:input_structure_text} demonstrates an example of an property textual and visual content that is prompted to the HotelMatch-LLM to being embedded.
\begin{figure}[ht]
    \centering
    \begin{mdframed}[backgroundcolor=verylightgray,roundcorner=10pt]
        \texttt{<image\_start>} \\
        \texttt{images\_embeddings} \\
        \texttt{<image\_end>} \\
        \texttt{country\_token} \\
        \texttt{city\_token} \\
        \texttt{property\_description}
    \end{mdframed}
    \caption{Representation of the model input structure. Images are represented as patch embeddings between the <image\_start> and <image\_end> tokens, followed by textual tokens for country, city, and property description.}
    \label{fig:input_structure_text}
\end{figure}
\end{document}